\documentstyle[12pt]{article}

\begin{document}

\begin{flushright}
DPUR/TH/65\\
March, 2020\\
\end{flushright}
\vspace{20pt}

\pagestyle{empty}
\baselineskip15pt

\begin{center}
{\large\bf  Emergence of Einstein Gravity from Weyl Gravity
\vskip 1mm }

\vspace{20mm}

Ichiro Oda\footnote{
           E-mail address:\ ioda@sci.u-ryukyu.ac.jp
                  }

\vspace{10mm}
           Department of Physics, Faculty of Science, University of the 
           Ryukyus,\\
           Nishihara, Okinawa 903-0213, Japan\\

\end{center}


\vspace{10mm}
\begin{abstract}

It is shown that in a quadratic gravity based on Weyl's conformal geometry, not only the Einstein-Hilbert 
action emerges but also a Weyl gauge field becomes massive in the Weyl gauge condition, $\tilde R = k$,  
for a Weyl gauge symmetry within the framework of the BRST formalism. We also consider a more general 
gravitational theory with a scalar field in the Weyl geometry and see that the Einstein-Hilbert action can be 
induced from spontaneous symmetry breakdown of the Weyl gauge symmetry. Thus, it turns out that 
Weyl's conformal gravity is quantum mechanically equivalent to Einstein's general relativity plus 
a massive Weyl gauge field, and the Weyl geometry is free from an infamous ``second clock problem'' 
in quantum regime.  

\end{abstract}

\newpage
\pagestyle{plain}
\pagenumbering{arabic}


\section{Introduction}

About one hundred years ago, shortly after the appearance of general relativity (GR) by Einstein, 
Weyl proposed a natural and profound generalization of the Riemann geometry \cite{Weyl}. This generalized
geometry essentially includes a vector gauge field called the Weyl gauge field in addition to a conventional 
metric tensor field, and is nowadays called the Weyl geometry. Weyl's idea is that since the Riemann 
geometry describes gravitation, a more general affine geometry might describe both gravitation 
and electromagnetism. Then, just as gravitation can be thought as being due to the path dependence 
of vectorial directions induced by the metric tensor field, electromagnetism could be considered 
as being due to the path dependence of vectorial magnitudes induced by the Weyl gauge field. 

Unfortunately for Weyl, Einstein immediately objected in a postscript of a Weyl's original paper \cite{Weyl} 
that in the Weyl gauge theory chemical elements with spectral lines of definite frequency could not exist
and the relative frequency of two neighbouring atoms of the same kind would be different in general,
depending on their past history, which is obviously in contradiction with observation.

With hindsight, in the Weyl gauge theory, there seem to be two closely related but distinct problems, 
one of which is that the Weyl gauge field cannot be identified with the electromagnetic potential, 
thereby nullifying the Weyl theory as a unified theory of gravitation and electromagnetism. The other problem 
can be paraphrased as the $\textit{second clock problem}$: Two clocks synchronized at the same space-time 
point display different clock rates after traveling along distinct paths and arriving at a common space-time 
point again in the presence of the Weyl gauge field \cite{Penrose}. 

With respect to the former problem, quantum mechanics came to the rescue of Weyl's gauge theory. 
What London noticed \cite{London} is that Weyl's idea should be used in a different manner: 
A scale factor multiplied under parallel transport of a vector might be changed by a phase factor involving 
$\hbar$, and it is attached to the quantum-mechanical wave function $\psi$ rather than the metric tensor 
field. Note that a pure phase shift in the wave function is not physically observable so Einstein's objection 
can be avoided by quantum mechanics. In this way, Weyl's attempt at combining gravitation with electromagnetism 
actually ended in failure, but the Weyl theory paved the way for the appearance of modern gauge theories \cite{O'Raifeartaigh}.

Nevertheless, the original Weyl gauge theory is still physically viable unless one insists on
identifying the Weyl gauge field with the electromagnetic gauge field since the electromagnetic field is described 
in terms of $U(1)$ gauge group whereas the Weyl gauge field is generated by a non-compact Abelian group.
In fact, the Weyl theory has been studied from various perspectives for a long time 
\cite{Omote}-\cite{Sanomiya}\footnote{See \cite{Scholz} for a historical review.} and it could be regarded
as one of the simplest gravitational theories with a non-compact gauge field. 

What then becomes of the latter problem, i.e., the second clock problem? This problem is closely related to 
the treatment of distance (or proper time) between the space-time point $x^\mu$ and its infinitesimal neigborhood 
$x^\mu + d x^\mu$. In the Riemann geometry, one can define a line element $d s^2 = g_{\mu\nu} d x^\mu d x^\nu$,
which is invariant under diffeomorphisms. If we use this line element even in the Weyl geometry, we encounter the
second clock problem. However, this line element is not invariant under the Weyl gauge transformation and cannot
have any definite physical meaning so it should be modified in such a way that a new line element is invariant
under the Weyl gauge transformation as well as diffeomorphisms. In fact, such a line element has been already proposed 
in Refs. \cite{Utiyama1, Freund, Utiyama2}. The idea is to introduce a scalar field $\phi (x)$ of weight $-1$ 
into the line element to compensate for the metric tensor of weight 2 and then construct a gauge invariant line element, 
$d \tilde s^2 = \phi^2 g_{\mu\nu} d x^\mu d x^\nu$. With this choice of the line element, we can automatically circumvent 
the infamous second clock problem that the rate of ``clocks'' depends on their ``history'', as will be also discussed 
in the last section. 

Notwithstanding this proposal for the gauge invariant line element, there still remains one question: What is
the relation between $d s$ and $d \tilde s$? We would like to respond to this question that quantum field theory would 
come the rescue of this issue in a similar way that quantum mechanics came to the rescue of Weyl's gauge theory. 
Recall that in quantum field theory, one must impose a gauge fixing condition from the beginning to quantize a theory.  
It is then natural to require that once the Weyl gauge invariance is fixed by a gauge fixing condition, and consequently
we have only diffeomorphisms as residual symmetries, the gauge invariant line element $d \tilde s$ should be reduced
to the line element $d s$ in the Riemann geometry since the gauge fixed theory is described by means of the 
Riemann geometry. Note that this requirement, in turn, gives us information as to the gauge fixing condition for
the Weyl gauge symmetry.

In this article, we wish to shed light on the relationship between Einstein's general relativity and the Weyl gauge
theory, in particular, a problem of how general relativity with a positive cosmological constant can be generated
from the Weyl gauge theory via the gauge fixing procedure of the Weyl gauge invariance in the BRST formalism.
This problem is of course not new since starting with an action of quadratic gravity, Weyl himself considered 
the relation to general relativity by taking what we call the Weyl gauge condition, $\tilde R = k$. We would like to 
extend the Weyl's study to the quantum regime on the basis of the BRST formalism. Moreover, we will deal with
a more general Weyl gauge theory involving a scalar field and work with a more general gauge condition,
$a \tilde R + b \phi^2 = k$. In this generalized theory, we find that the Higgs potential is naturally generated,
thereby triggering spontaneous symmetry breakdown of the Weyl gauge invariance. In the process of obtaining
these results, we find that the so-called "the second clock problem" is naturally solved within the framework of 
quantum field theories. 

We close this section with an overview of this article.
In Section 2, we briefly review the Weyl geometry and the Weyl gauge theory, details of which can be
found in other literatures \cite{Weyl, O'Raifeartaigh}.  
In Section 3, we present a more general action of Weyl's quadratic gravity including a scalar field, 
and derive field equations from it.  In Section 4, we present a gauge invariant line element and take a Weyl
gauge condition for the Weyl gauge transformation. It is shown that by a suitable choice of a gauge parameter
the Einstein-Hilbert action can be derived from the action of Weyl's quadratic gravity.
Moreover, it is shown that in the Einstein or unitary gauge the number of physical degrees of freedom
is unchanged before and after spontaneous symmetry breakdown (SSB).
In Section 5, we work with a general action and take a more general gauge condition where we see that
the SSB of the Weyl gauge symmetry induces the Einstein-Hilbert action and a massive Weyl gauge field at the
same time. Section 6 is devoted to the conclusion.

\section{Review of Weyl conformal geometry} 
We briefly review the basic concepts and definitions of the Weyl conformal geometry.\footnote{See
also Refs. \cite{Adler, Smolin, Cesare} for a concise introduction of the Weyl geometry.} 
In the Weyl geometry, the Weyl gauge transformation, which is the sum of a local scale transformation for 
a generic field $\Phi (x)$ and a gauge transformation for the Weyl gauge field $S_\mu(x)$, is defined as
\begin{eqnarray}
\Phi (x) \rightarrow \Phi^\prime (x) = e^{w \Lambda(x)} \Phi (x), \qquad
S_\mu (x) \rightarrow S^\prime_\mu (x) = S_\mu (x) - \frac{1}{f} \partial_\mu \Lambda (x),
\label{Weyl transf}
\end{eqnarray}
where $w$ is called the ``Weyl weight'', or simply ``weight'' henceforth, $f$ is the coupling constant 
for the non-compact Abelian gauge group, and $\Lambda(x)$ is a local parameter for the Weyl transformation. 
The Weyl gauge transformation for various fields is explicitly given by
\begin{eqnarray}
g_{\mu\nu} (x) &\rightarrow& g_{\mu\nu}^\prime (x) = e^{2 \Lambda(x)} g_{\mu\nu}(x), \qquad
\phi (x) \rightarrow \phi^\prime (x) = e^{- \Lambda(x)} \phi (x),  \nonumber\\
\psi (x) &\rightarrow& \psi^\prime (x) = e^{- \frac{3}{2} \Lambda(x)} \psi (x), \qquad
A_\mu (x) \rightarrow A^\prime_\mu (x) = A_\mu (x),
\label{Weyl transf 2}
\end{eqnarray}
where $g_{\mu\nu} (x)$, $\phi (x)$, $\psi (x)$ and $A_\mu (x)$ are the metric tensor, scalar, spinor,
and electromagnetic gauge fields, respectively. The covariant derivative $D_\mu$ for the Weyl gauge
transformation for a generic field $\Phi (x)$ of weight $w$ is defined as
\begin{eqnarray}
D_\mu \Phi \equiv \partial_\mu \Phi + w f S_\mu \Phi,
\label{W-cov-deriv}
\end{eqnarray}
which transforms covariantly under the Weyl transformation:
\begin{eqnarray}
D_\mu \Phi \rightarrow (D_\mu \Phi)^\prime = e^{w \Lambda(x)} D_\mu \Phi.
\label{S-cov-transf}
\end{eqnarray}

The Weyl geometry is defined as a geometry with a real symmetric metric tensor $g_{\mu\nu}
(= g_{\nu\mu})$ and a symmetric connection $\tilde \Gamma^\lambda_{\mu\nu} (= \tilde \Gamma^\lambda_{\nu\mu})$ 
which is defined as\footnote{We often use the tilde characters to express quantities belonging to the Weyl geometry.}
\begin{eqnarray}
\tilde \Gamma^\lambda_{\mu\nu} &=& \frac{1}{2} g^{\lambda\rho} \left( D_\mu g_{\nu\rho} + D_\nu g_{\mu\rho}
- D_\rho g_{\mu\nu} \right)
\nonumber\\
&=& \Gamma^\lambda_{\mu\nu} + f \left( S_\mu \delta^\lambda_\nu + S_\nu \delta^\lambda_\mu 
- S^\lambda g_{\mu\nu} \right),
\label{W-connection}
\end{eqnarray}
where 
\begin{eqnarray}
\Gamma^\lambda_{\mu\nu} \equiv \frac{1}{2} g^{\lambda\rho} \left( \partial_\mu g_{\nu\rho} 
+ \partial_\nu g_{\mu\rho} - \partial_\rho g_{\mu\nu} \right),
\label{Affine connection}
\end{eqnarray}
is the Christoffel symbol in the Riemann geometry. The most important difference between the Riemann geometry 
and the Weyl one lies in the fact that in the Riemann geometry the metric condition is satisfied 
\begin{eqnarray}
\nabla_\lambda g_{\mu\nu} \equiv \partial_\lambda g_{\mu\nu} - \Gamma^\rho_{\lambda\mu} 
g_{\rho\nu} - \Gamma^\rho_{\lambda\nu} g_{\mu\rho} = 0, 
\label{Metric cond}
\end{eqnarray}
while in the Weyl geometry we have
\begin{eqnarray}
\tilde \nabla_\lambda g_{\mu\nu} \equiv \partial_\lambda g_{\mu\nu} - \tilde \Gamma^\rho_{\lambda\mu} 
g_{\rho\nu} - \tilde \Gamma^\rho_{\lambda\nu} g_{\mu\rho}
= - 2 f S_\lambda g_{\mu\nu},
\label{W-metric cond}
\end{eqnarray}
where $\nabla_\mu$ and $\tilde \nabla_\mu$ are covariant derivatives for diffeomorphisms in the Riemann 
and Weyl geometries, respectively. Since the metric condition (\ref{Metric cond}) implies that both length 
and angle are preserved under parallel transport, Eq. (\ref{W-metric cond}) shows that only angle, 
but not length, is preserved by the Weyl connection.

The general covariant derivative for both diffeomorphisms and the Weyl gauge transformation, for instance,
for a covariant vector of weight $w$, is defined as
\begin{eqnarray}
{\cal D}_\mu V_\nu &\equiv& D_\mu V_\nu - \tilde \Gamma^\rho_{\mu\nu} V_\rho  \nonumber\\
&=& \tilde \nabla_\mu V_\nu + w f S_\mu V_\nu \nonumber\\
&=& \nabla_\mu V_\nu + w f S_\mu V_\nu - f ( S_\mu \delta^\rho _\nu + S_\nu \delta^\rho _\mu
- S^\rho g_{\mu\nu} ) V_\rho \nonumber\\
&=& \partial_\mu V_\nu + w f S_\mu V_\nu - \Gamma^\rho_{\mu\nu} V_\rho
- f ( S_\mu \delta^\rho _\nu + S_\nu \delta^\rho _\mu - S^\rho g_{\mu\nu} ) V_\rho.
\label{Gen-cov-deriv}
\end{eqnarray}
One can verify that using the general covariant derivative, the following metric condition is 
satisfied:
\begin{eqnarray}
{\cal D}_\lambda g_{\mu\nu} = 0.
\label{Gen-metric cond}
\end{eqnarray}
 Moreover, under the Weyl gauge transformation the general covariant derivative for a generic field 
 $\Phi$ of weight $w$ transforms in a covariant manner as desired:
\begin{eqnarray}
{\cal D}_\mu \Phi \rightarrow ({\cal D}_\mu \Phi)^\prime = e^{w \Lambda(x)} {\cal D}_\mu \Phi,
\label{Gen-cov-transf}
\end{eqnarray}
because the Weyl connection is invariant under the Weyl gauge transformation, i.e., 
$\tilde \Gamma^{\prime \rho}_{\mu\nu} = \tilde \Gamma^\rho_{\mu\nu}$.

As in the Riemann geometry, in the Weyl geometry one can also construct a Weyl invariant curvature
tensor $\tilde R_{\mu\nu\rho} \, ^\sigma$ via a commutator of the covariant 
derivative $\tilde \nabla_\mu$\footnote{In this article, we make use of the conventions and notation 
for the Riemann tensors and the metric signature in the Wald textbook \cite{Wald}, and in particular 
our sign convention is $\eta_{\mu\nu} = diag ( - 1, 1, 1, 1)$. }
\begin{eqnarray}
[ \tilde \nabla_\mu, \tilde \nabla_\nu ] V_\rho = \tilde R_{\mu\nu\rho} \, ^\sigma V_\sigma.
\label{Commutator}
\end{eqnarray}
Calculating this commutator, one finds that
\begin{eqnarray}
\tilde R_{\mu\nu\rho} \, ^\sigma &=& \partial_\nu \tilde \Gamma^\sigma_{\mu\rho} 
- \partial_\mu \tilde \Gamma^\sigma_{\nu\rho} + \tilde \Gamma^\alpha_{\mu\rho} \tilde \Gamma^\sigma_{\alpha\nu} 
- \tilde \Gamma^\alpha_{\nu\rho} \tilde \Gamma^\sigma_{\alpha\mu}
\nonumber\\
&=& R_{\mu\nu\rho} \, ^\sigma + 2 f \left( \delta^\sigma_{[\mu} \nabla_{\nu]} S_\rho 
- \delta^\sigma_\rho \nabla_{[\mu} S_{\nu]} - g_{\rho [\mu} \nabla_{\nu]} S^\sigma \right)
\nonumber\\
&+& 2 f^2 \left( S_{[\mu} \delta^\sigma_{\nu]} S_\rho - S_{[\mu} g_{\nu]\rho} S^\sigma
+ \delta^\sigma_{[\mu} g_{\nu]\rho} S_\alpha S^\alpha \right),
\label{W-curv-tensor}
\end{eqnarray}
where $R_{\mu\nu\rho} \, ^\sigma$ is the curvature tensor in the Riemann geometry, and
we have defined the antisymmetrization by the square bracket, e.g., $A_{[\mu} B_{\nu]} \equiv 
\frac{1}{2} ( A_\mu B_\nu - A_\nu B_\mu )$. Then, it is straightforward to prove the following identities:
\begin{eqnarray}
\tilde R_{\mu\nu\rho} \, ^\sigma = - \tilde R_{\nu\mu\rho} \, ^\sigma,  \qquad
\tilde R_{[\mu\nu\rho]} \, ^\sigma = 0, \qquad
\tilde \nabla_{[\lambda} \tilde R_{\mu\nu]\rho} \, ^\sigma = 0.
\label{W-curv-identity}
\end{eqnarray}
The curvature tensor $\tilde R_{\mu\nu\rho} \, ^\sigma$ has $26$ independent components,
twenty of which are possessed by $R_{\mu\nu\rho} \, ^\sigma$ and six by the Weyl
invariant field strength $H_{\mu\nu} \equiv \partial_\mu S_\nu - \partial_\nu S_\mu$.

From $\tilde R_{\mu\nu\rho} \, ^\sigma$ one can define a Weyl invariant 
Ricci tensor:
\begin{eqnarray}
\tilde R_{\mu\nu} &\equiv& \tilde R_{\mu\rho\nu} \, ^\rho
\nonumber\\
&=& R_{\mu\nu} + f \left( - 2 \nabla_\mu S_\nu - H_{\mu\nu} - g_{\mu\nu} \nabla_{\alpha} S^\alpha \right)
\nonumber\\
&+& 2 f^2 \left( S_\mu S_\nu - g_{\mu\nu} S_\alpha S^\alpha \right).
\label{W-Ricci-tensor}
\end{eqnarray}
Let us note that 
\begin{eqnarray}
\tilde R_{[\mu\nu]} \equiv \frac{1}{2} ( \tilde R_{\mu\nu} - \tilde R_{\nu\mu} ) = - 2 f H_{\mu\nu}.
\label{W-Ricci-tensor 2}
\end{eqnarray}
Similarly, one can define a not Weyl invariant but Weyl covariant scalar curvature:
\begin{eqnarray}
\tilde R \equiv g^{\mu\nu} \tilde R_{\mu\nu} 
= R - 6 f \nabla_\mu S^\mu - 6 f^2 S_\mu S^\mu.
\label{W-scalar-curv}
\end{eqnarray}
One finds that under the Weyl gauge transformation, $\tilde R \rightarrow \tilde R^\prime = e^{- 2 \Lambda(x)}
\tilde R$ while $\tilde \Gamma^\lambda_{\mu\nu}, \tilde R_{\mu\nu\rho} \, ^\sigma$ and $\tilde R_{\mu\nu}$
are all invariant.

Even in the Weyl geometry, it is possible to write out a generalization of the Gauss-Bonnet topological 
invariant which can be described as
\begin{eqnarray}
I_{GB} &\equiv& \int d^4 x \sqrt{-g} \, \epsilon^{\mu\nu\rho\sigma} \epsilon_{\alpha\beta\gamma\delta} \, 
\tilde R_{\mu\nu} \, ^{\alpha\beta} \tilde R_{\rho\sigma} \, ^{\gamma\delta}    
\nonumber\\
&=& - 2 \int d^4 x \sqrt{-g} \,  \left( \tilde R_{\mu\nu\rho\sigma} \tilde R^{\rho\sigma\mu\nu} 
- 4 \tilde R_{\mu\nu} \tilde R^{\nu\mu} + \tilde R^2 - 12 f^2 H_{\mu\nu} H^{\mu\nu} \right)
\nonumber\\
&=& - 2 \int d^4 x \sqrt{-g} \,  \left( R_{\mu\nu\rho\sigma} R^{\mu\nu\rho\sigma} 
- 4 R_{\mu\nu} R^{\mu\nu} + R^2 \right).   
\label{GB}
\end{eqnarray}

We close this section by discussing a spinor field as an example of matter fields in the Weyl
geometry \cite{Shirafuji, Hayashi}.  As is well known, to describe a spinor field it is necessary to introduce 
the vierbein $e^a _\mu$, which is defined as
\begin{eqnarray}
g_{\mu\nu} = \eta_{ab} e^a _\mu e^b _\nu,
\label{Vierbein}
\end{eqnarray}
where $a, b, \cdots$ are local Lorentz indices taking $0, 1, 2, 3$ and $\eta_{ab} = diag ( - 1, 1, 1, 1)$.

Now the metric condition (\ref{Gen-metric cond}) takes the form 
\begin{eqnarray}
{\cal D}_\mu e^a _\nu \equiv D_\mu e^a _\nu + \tilde \omega^a \, _{b \mu} e^b _\nu 
- \tilde \Gamma^\rho_{\mu\nu} e^a _\rho = 0,
\label{Gen-vierbein cond}
\end{eqnarray}
where the general covariant derivative is extended to include the local Lorentz transformation whose
gauge connection is the spin connection $\tilde \omega^a \, _{b \mu}$ of weight $0$ in the Weyl
geometry, and $D_\mu e^a _\nu = \partial_\mu e^a _\nu + f S_\mu e^a _\nu$ since the vierbein 
$e^a _\mu$ has weight $1$. Solving the metric condition (\ref{Gen-vierbein cond}) leads to the
expression of the spin connection in the Weyl geometry
\begin{eqnarray}
\tilde \omega_{a b \mu} = \omega_{a b \mu} + f e^c _\mu ( \eta_{ac} S_b - \eta_{bc} S_a ),
\label{spin connection}
\end{eqnarray}
where $\omega_{a b \mu}$ is the spin connection in the Riemann geometry and we have defined 
$S_a \equiv e^\mu _a S_\mu$. Then, the general covariant 
derivative for a spinor field $\Psi$ of weight $- \frac{3}{2}$ reads
\begin{eqnarray}
{\cal D}_\mu \Psi = D_\mu \Psi + \frac{i}{2} \tilde \omega_{a b \mu} S^{a b}  \Psi,
\label{spinor CD}
\end{eqnarray}
where $D_\mu \Psi = \partial_\mu \Psi - \frac{3}{2} f S_\mu \Psi$ and the Lorentz generator $S^{a b}$
for a spinor field is defined as $S^{a b} = \frac{i}{4} [ \gamma^a, \gamma^b ]$. Here we define the gamma
matrices to satisfy the Clifford algebra $\{ \gamma^a, \gamma^b \} = - 2 \eta^{ab}$.
Since the spin connection $\tilde \omega^a \, _{b \mu}$ has weight $0$, the covariant
derivative ${\cal D}_\mu \Psi$  transforms covariantly under the Weyl gauge transformation
\begin{eqnarray}
{\cal D}_\mu \Psi \rightarrow ( {\cal D}_\mu \Psi )^\prime = e^{- \frac{3}{2} \Lambda(x)} 
{\cal D}_\mu \Psi.
\label{spinor covariance}
\end{eqnarray}

Then, the Lagrangian density for a massless Dirac spinor field is of form
\begin{eqnarray}
{\cal L} = \frac{i}{2} e \ e^\mu _a ( \bar \Psi \gamma^a {\cal D}_\mu \Psi 
- {\cal D}_\mu \bar \Psi \gamma^a \Psi ),
\label{spinor Lag}
\end{eqnarray}
where $e \equiv \sqrt{-g}, \bar \Psi \equiv \Psi^\dagger \gamma^0$, and ${\cal D}_\mu \bar \Psi$
is given by
\begin{eqnarray}
{\cal D}_\mu \bar \Psi = D_\mu \bar \Psi - \bar \Psi \frac{i}{2} \tilde \omega_{a b \mu} S^{a b}.
\label{spinor CD2}
\end{eqnarray}
Inserting Eqs. (\ref{spinor CD}) and (\ref{spinor CD2}) to the Lagrangian density (\ref{spinor Lag}), 
we find that  
\begin{eqnarray}
{\cal L} &=& \frac{i}{2} e \Bigl[ e^\mu _a  \left( \bar \Psi \gamma^a \partial_\mu \Psi 
- \partial_\mu \bar \Psi \gamma^a \Psi + \frac{i}{2} \omega_{b c \mu} \bar \Psi \{ \gamma^a,
S^{bc} \} \Psi \right)
\nonumber\\
&+& \frac{i}{2} f ( \eta_{ab} S_c - \eta_{ac} S_b ) \bar \Psi \{ \gamma^a, S^{bc} \} \Psi \Bigr].
\label{spinor Lag2}
\end{eqnarray}
The last term identically vanishes owing to the relation 
\begin{eqnarray}
\{ \gamma^a, S^{bc} \} = - \varepsilon^{abcd} \gamma_5 \gamma_d,
\label{gamma rel}
\end{eqnarray}
where we have defined as $\gamma_5 = i \gamma^0 \gamma^1 \gamma^2 \gamma^3$ and 
$\varepsilon^{0123} = +1$. Thus, as is well known, the Weyl gauge field $S_\mu$ does not couple minimally 
to a spinor field $\Psi$.  Technically speaking, it is the absence of imaginary unit $i$ in the 
covariant derivative $D_\mu \Psi = \partial_\mu \Psi - \frac{3}{2} f S_\mu \Psi$ that induced this
decoupling of the Weyl gauge field from the spinor field. Without the imaginary unit, the terms including 
the Weyl gauge field cancel out each other in Eq. (\ref{spinor Lag}). In a similar manner, we can prove that 
the Weyl gauge field does not couple to a gauge field either such as the electromagnetic potential $A_\mu$. 
On the other hand, the Weyl gauge field can couple to a scalar field such as the Higgs field as well as a graviton. 
In such a situation, we cannot help identifying the Weyl gauge field with an elementary particle that constitutes 
dark matter. It seems that the Weyl gauge theory was rejected as a unified theory of gravitation and electromagnetism
but it has revived as a geometrical theory which predicts the existence of dark matter.

\section{Classical theory}
In this section, we wish to present the most general action of Weyl's gauge theory involving a scalar field
and derive field equations.

One of interesting features in the Weyl gauge theory is that if only the metric tensor is allowed to use for 
the construction of a gravitational action, the action invariant under the Weyl transformation must be 
of form of quadratic gravity, but is neither of the Einstein-Hilbert term nor of higher derivative terms such as
$\tilde R^3$. Using the topological invariant (\ref{GB}), 
one can write down a general action of quadratic gravity, which is invariant under the Weyl transformation, as follows:
\begin{eqnarray}
S_{QG} = \int d^4 x \sqrt{-g} \left( \frac{\xi_0}{2}  \tilde C_{\mu\nu\rho\sigma} \tilde C^{\mu\nu\rho\sigma} 
+ \frac{\xi_1}{2} \tilde R^2 \right),
\label{QG}
\end{eqnarray}
where $\xi_0$ and $\xi_1$ are dimensionless coupling constants. And a generalization of the conformal tensor, 
$\tilde C_{\mu\nu\rho\sigma}$, in the Weyl geometry is defined as in $C_{\mu\nu\rho\sigma}$ in the Riemann geometry:
\begin{eqnarray}
\tilde C_{\mu\nu\rho\sigma} &\equiv& \tilde R_{\mu\nu\rho\sigma} - \frac{1}{2} ( g_{\mu\rho} \tilde R_{\nu\sigma}
+ g_{\nu\sigma} \tilde R_{\mu\rho} -  g_{\mu\sigma} \tilde R_{\nu\rho} - g_{\nu\rho} \tilde R_{\mu\sigma} )
\nonumber\\
&+& \frac{1}{6} ( g_{\mu\rho} g_{\nu\sigma} - g_{\mu\sigma} g_{\nu\rho} ) \tilde R
\nonumber\\
&=& C_{\mu\nu\rho\sigma} + f \biggl[ - g_{\rho\sigma} H_{\mu\nu} + \frac{1}{2} ( g_{\mu\rho} H_{\nu\sigma}
+  g_{\nu\sigma} H_{\mu\rho} 
\nonumber\\
&-& g_{\mu\sigma} H_{\nu\rho} - g_{\nu\rho} H_{\mu\sigma} ) \biggr].
\label{Conformal tensor}
\end{eqnarray}
This conformal tensor in the Weyl geometry has the following properties:
\begin{eqnarray}
\tilde C_{\mu\nu\rho\sigma} = - \tilde C_{\nu\mu\rho\sigma}, \qquad 
\tilde C_{\mu\nu\rho} \, ^\nu = 0, \qquad 
\tilde C_{\mu\nu\rho} \, ^\rho = - 4 f H_{\mu\nu}.
\label{Conformal tensor 2}
\end{eqnarray}
 
Now, to this quadratic action let us add a kinetic action for the Weyl gauge field and an action including
a scalar field $\phi$ of weight $-1$
\begin{eqnarray}
S &=& \int d^4 x \sqrt{-g} \biggl( \frac{\xi_0}{2}  \tilde C_{\mu\nu\rho\sigma} \tilde C^{\mu\nu\rho\sigma} 
+ \frac{\xi_1}{2} \tilde R^2 - \frac{1}{4} H_{\mu\nu} H^{\mu\nu} 
\nonumber\\
&+& \frac{\xi_2}{2} \phi^2 \tilde R - \frac{1}{2} g^{\mu\nu} D_\mu \phi D_\nu \phi - \frac{\lambda}{4} \phi^4 \biggr),
\label{Action}
\end{eqnarray}
where $\xi_2$ and $\lambda$ are also dimensionless coupling constants and $D_\mu \phi \equiv \partial_\mu \phi
- f S_\mu \phi$. Provided that only the metric tensor and a scalar field are allowed to use, 
the action (\ref{Action}) is the most general and renormalizable action in the Weyl geometry, 
but the first term violates the unitarity \cite{Stelle}, so we shall drop this term by taking $\xi_0 = 0$ from now on. 
The action which we consider in this article, therefore, takes the form  
\begin{eqnarray}
S = \int d^4 x \sqrt{-g} \biggl( \frac{\xi_1}{2} \tilde R^2 - \frac{1}{4} H_{\mu\nu} H^{\mu\nu} 
+ \frac{\xi_2}{2} \phi^2 \tilde R - \frac{1}{2} g^{\mu\nu} D_\mu \phi D_\nu \phi - \frac{\lambda}{4} \phi^4 \biggr).
\label{Action2}
\end{eqnarray}

From the action (\ref{Action2}), let us derive field equations for the Weyl gauge field $S_\mu$, the scalar 
field $\phi$ and the gravitational field $g_{\mu\nu}$. The field equation for the Weyl gauge field is given by
\begin{eqnarray}
\nabla^\nu H_{\mu\nu} = 6 f D_\mu ( \xi_1 \tilde R + \frac{6 \xi_2 + 1}{12} \phi^2 ).
\label{S-eq of motion}
\end{eqnarray}
The identity $\nabla^\mu \nabla^\nu H_{\mu\nu} = 0$ gives rise to a relation
\begin{eqnarray}
\nabla^\mu D_\mu ( \xi_1 \tilde R + \frac{6 \xi_2 + 1}{12} \phi^2 ) = 0.
\label{S-iden}
\end{eqnarray}

It is easy to derive the field equation for the scalar field whose result is written as
\begin{eqnarray}
\frac{6 \xi_2 + 1}{6} \phi \tilde R = \frac{1}{6} \phi R - \Box \phi + \lambda \phi^3.
\label{phi-eq of motion}
\end{eqnarray}
Note that  in Eqs. (\ref{S-eq of motion}), (\ref{S-iden}) and (\ref{phi-eq of motion}) a specific value 
$\xi_2 = - \frac{1}{6}$ leads to some simplication of field equations as previously mentioned 
by Dirac \cite{Dirac}.

Finally, the field equation for the gravitational field gives rise to a little lengthy expression
\begin{eqnarray}
&{}& \xi_1 \tilde R ( \tilde R_{(\mu\nu)} - \frac{1}{4} g_{\mu\nu} \tilde R )
+ \frac{\xi_2}{2} \phi^2 ( \tilde R_{(\mu\nu)} - \frac{1}{2} g_{\mu\nu} \tilde R )
\nonumber\\
&=& \frac{1}{2} ( H_{\mu\alpha} H_\nu \,^\alpha - \frac{1}{4} g_{\mu\nu} H_{\alpha\beta} H^{\alpha\beta} ) 
+ \frac{1}{2} ( D_\mu \phi D_\nu \phi - \frac{1}{2} g_{\mu\nu} D_\alpha \phi D^\alpha \phi )
- \frac{\lambda}{8} g_{\mu\nu} \phi^4
\nonumber\\
&+& 4 f K \left[ \nabla_\mu S_\nu - \frac{1}{4} g_{\mu\nu} \nabla_\alpha S^\alpha
+ 2 f ( S_\mu S_\nu - \frac{1}{4} g_{\mu\nu} S_\alpha S^\alpha ) \right]
\nonumber\\
&+& 6 f \left[ \frac{1}{2} g_{\mu\nu} \nabla_\alpha ( K S^\alpha ) - \nabla_{(\mu} ( S_{\nu)} K ) \right]
+ ( \nabla_\mu \nabla_\nu - g_{\mu\nu} \Box ) K,
\label{g-eq of motion}
\end{eqnarray}
where we have defined $K \equiv \xi_1 \tilde R + \frac{\xi_2}{2} \phi^2$ and the symmetrization by
the round bracket, e.g., $A_{(\mu} B_{\nu)} \equiv \frac{1}{2} ( A_\mu B_\nu + A_\nu B_\mu)$. 
Taking the trace of this equation, we have
\begin{eqnarray}
- \frac{\xi_2}{2} \phi^2 \tilde R = - \frac{1}{2} D_\alpha \phi D^\alpha \phi - \frac{\lambda}{2} \phi^4
+ 6 f \nabla_\alpha ( K S^\alpha ) - 3 \Box K.
\label{g-eq of motion2}
\end{eqnarray}
Here we make use of Eq. (\ref{S-iden}), which can be rewritten as
\begin{eqnarray}
\Box K = - \frac{1}{12} \Box \phi^2 + 2 f \nabla_\alpha \left[ ( \xi_1 \tilde R + \frac{6 \xi_2 + 1}{12} \phi^2
S^\alpha ) \right].
\label{S-iden2}
\end{eqnarray}
Substituting Eq.  (\ref{S-iden2}) into  (\ref{g-eq of motion2}), we can obtain (\ref{phi-eq of motion}). Thus,
the field equation for the scalar field $\phi$ is not independent but derived from the other field equations
due to the Weyl gauge symmetry.

\section{Weyl gauge} 

In order to clarify quantum aspects of a general gauge theory, we must impose a gauge fixing
condition from the beginning. In this sense, the gauge fixing conditions play an essential role in 
quantum field theory and their importance should not be underestimated. For instance,
in order to show that an Utiyama's incorrect claim, that is, Weyl's gauge field has a negative energy 
\cite{Utiyama1, Utiyama2}, Hayashi and Kugo quantized the Utiyama's action by imposing the ``Einstein gauge 
condition'' $\phi (x) = 1$ for the Weyl gauge transformation in a manifestly covariant way, and then proved 
that the gauge field has a positive energy density, thus having a normal behaviour \cite{Hayashi}.
  
Thus far, we are familiar with four kinds of gauge fixing conditions in Weyl gravity, thereby making us easy 
to understand the connection between the Weyl gravity and Einstein's theory of gravitation. 
The first gauge fixing condition is called the ``Weyl gauge'' putting Weyl's scalar curvature to 
a constant, i.e., $\tilde R = k$ \cite{Weyl}, which will be discussed in detail in this section. 
The second gauge fixing condition is called the ``Einstein gauge'', which sets a scalar field called
the measure field by Utiyama to a constant. This gauge fixing will be later discussed in this section as well
since the Einstein gauge somewhat resembles the ``unitary gauge'' in the Higgs model with spontaneous symmetry
breakdown and shows the particle content in a manifest manner.
The third gauge fixing condition, which we will name a ``general gauge'', is a linear combination of the Einstein gauge 
and the Weyl gauge \cite{Freund}, and will be argued in the next section.
The final gauge condition was the ``Lorenz gauge'' $\nabla_\mu S^\mu = 0$ which was used in evaluating an
effective action of a quadratic gravity in the Weyl geometry \cite{Oda2}.

Let us start with a quadratic gravity with the curvature squared and the kinetic term for the Weyl gauge field 
whose Lagrangian density is given by
\begin{eqnarray}
{\cal{L}} = \sqrt{-g} \left( \frac{\xi_1}{2} \tilde R^2 - \frac{1}{4} H_{\mu\nu} H^{\mu\nu} \right),
\label{QG-Lag1}
\end{eqnarray}
where a coupling constant $\xi_1$ is taken to be positive in order to avoid a ghost.
The field equations for the Weyl gauge field $S_\mu$ and the metric field $g_{\mu\nu}$ can be obtained
by putting $\phi = 0$ in the field equations derived in the previous section. 
The field equation for the Weyl gauge field reads
\begin{eqnarray}
\nabla^\nu H_{\mu\nu} = 6 \xi_1 f D_\mu \tilde R.
\label{QG-S-eq of motion}
\end{eqnarray}
The identity $\nabla^\mu \nabla^\nu H_{\mu\nu} = 0$ then yields a relation
\begin{eqnarray}
\Box \tilde R = 2 f \nabla_\mu ( S^\mu \tilde R ).
\label{QG-S-iden}
\end{eqnarray}
Moreover, the field equation for the metric tensor field leads to
\begin{eqnarray}
&{}& \xi_1 \tilde R ( \tilde R_{(\mu\nu)} - \frac{1}{4} g_{\mu\nu} \tilde R )
\nonumber\\
&=& \frac{1}{2} ( H_{\mu\alpha} H_\nu \,^\alpha - \frac{1}{4} g_{\mu\nu} H_{\alpha\beta} H^{\alpha\beta} ) 
+ 4 \xi_1 f \tilde R \biggl[ \nabla_\mu S_\nu - \frac{1}{4} g_{\mu\nu} \nabla_\alpha S^\alpha
\nonumber\\
&+& 2 f ( S_\mu S_\nu - \frac{1}{4} g_{\mu\nu} S_\alpha S^\alpha ) \biggr]
+ 6 \xi_1 f \left[ \frac{1}{2} g_{\mu\nu} \nabla_\alpha ( S^\alpha \tilde R) 
- \nabla_{(\mu} ( S_{\nu)} \tilde R ) \right]
\nonumber\\
&+& \xi_1 ( \nabla_\mu \nabla_\nu - g_{\mu\nu} \Box ) \tilde R.
\label{QG-g-eq of motion}
\end{eqnarray}
Taking the trace of this equation gives us the same equation as in Eq. (\ref{QG-S-iden}).

Next, we wish to consider a gauge invariant line element in the sense that it is invariant under both 
diffeomorphisms and the Weyl gauge transformation. It is of interest to see that there exists almost 
a unique candidate and one can construct such a line element without introducing a scalar field:
\begin{eqnarray}
d \tilde s^2 = \tilde R g_{\mu\nu} d x^\mu d x^\nu \equiv \tilde R d s^2.
\label{line-element1}
\end{eqnarray}
Since the scalar curavature $\tilde R$ and the metric tensor $g_{\mu\nu}$ have weight $-2$ and $2$, 
respectively, a composite operator $\tilde R g_{\mu\nu}$ has weight $0$ so it is invariant under the Weyl
gauge transformation. Provided that the choice of a gauge condition makes the gauge invariant line element 
$d \tilde s$ coincide with the conventional line element $d s$ in the Riemann geometry (up to an overall constant),
we are invited to choose the Weyl gauge condition
\begin{eqnarray}
\tilde R = k,
\label{W-gauge}
\end{eqnarray}
where $k$ is a certain constant. Actually, with this gauge choice, the gauge invariant line element $d \tilde s$
reduces to that in the Riemann geometry as seen in Eq. (\ref{line-element1}).\footnote{If we want a complete
agreement between $d \tilde s$ and $d s$, we can change the definition of $d \tilde s^2$ like $d \tilde s^2 
= \frac{1}{k} \tilde R g_{\mu\nu} d x^\mu d x^\nu$.} 

Before delving into a quantum theory, we would like to examine whether or not the Weyl gauge condition (\ref{W-gauge})
allows a flat Minkowski metric $g_{\mu\nu} = \eta_{\mu\nu}$ as a classical solution. In particular, there might be
a possibility that we could not find a Lorentz covariant solution for the Weyl gauge field. To show that there is
indeed a Lorentz covariant solution, let us substitute the flat Minkowski metric into Eq. (\ref{W-gauge}) whose
result is rewritten as
\begin{eqnarray}
- 6 f \partial_\mu S^\mu - 6 f^2 S_\mu S^\mu = k.
\label{W-gauge-sol1}
\end{eqnarray}
As a Lorentz covariant solution, let us consider a pure gauge, $S_\mu = \partial_\mu \omega$ where 
$\omega$ is a scalar field. Inserting this pure gauge solution to Eq. (\ref{W-gauge-sol1}) produces an equation
\begin{eqnarray}
- 6 f \Box \omega - 6 f^2 (\partial_\mu \omega)^2  = k,
\label{W-gauge-sol2}
\end{eqnarray}
with being $\Box \equiv \eta^{\mu\nu} \partial_\mu \partial_\nu$. Assuming $\omega = c p_\mu x^\mu$
where $c$ is a constant and $p_\mu$ is a four dimensional momentum, Eq. (\ref{W-gauge-sol2})
leads to a relation
\begin{eqnarray}
p_\mu ^2 = - \frac{k}{6 f^2 c^2} \equiv - m^2,
\label{W-gauge-sol3}
\end{eqnarray}
where we have put $k = 6 ( f c m )^2$, which holds only for a positive $k$. Actually, as will be seen later
in Eq. (\ref{k-choice}), $k$ must be a positive constant. Thus, as long as $p_\mu$ satisfies the
``mass-shell condition'' (\ref{W-gauge-sol3}), there is a Lorentz covariant solution for the Weyl
gauge field, by which the flat Minkowski metric is a classical solution to the Weyl gauge 
condition (\ref{W-gauge}).

Next, in order to move to a quantum theory, we rely on the BRST formalism. The BRST formalism can be 
made by two steps: The first step is to gauge-fix the Weyl gauge symmetry and the second step is to 
do diffeomorphisms. Since each BRST charge commutes, we will treat with only the BRST formalism 
for the Weyl gauge symmetry. The construction of the BRST formalism for diffeomorphisms can be 
done in a similar way to Ref. \cite{Nakanishi}.

The BRST transformation for the Weyl gauge symmetry takes the form  
\begin{eqnarray}
\delta_B \sqrt{- g} &=& 4 c \sqrt{- g}, \qquad
\delta_B \tilde R = - 2 c \tilde R, \nonumber\\
\delta_B \bar c &=& i B, \qquad
\delta_B B = \delta_B c = 0,
\label{BRST1}
\end{eqnarray}
where $c, \bar c$ and $B$ are a ghost, an anti-ghost and the Nakanishi-Lautrup field, respectively. 
Then, the Lagrangian density for the gauge fixing term and the FP ghost term is given by \cite{Kugo}
\begin{eqnarray}
{\cal{L}}_{GF + FP} &=& - i \delta_B \left[ \sqrt{- g} \bar c ( \tilde R - k + \frac{1}{2} \alpha B ) \right]
\nonumber\\
&=& \sqrt{- g} \left[  B_* ( \tilde R - k ) + \frac{1}{2} \alpha B_* ^2 - 2 i k \bar c c ) \right],
\label{GF+FP1}
\end{eqnarray}
where $\alpha$ is a gauge parameter and we have defined $B_* \equiv B + 2 i \bar c c$.
After performing the integration over $B_*, c, \bar c$, we have
\begin{eqnarray}
{\cal{L}}_{GF + FP} = - \frac{1}{2 \alpha} \sqrt{- g} ( \tilde R - k )^2 - i \hbar \delta^4 (0) \log \sqrt{- g(x)}.
\label{GF+FP2}
\end{eqnarray}
The last term comes from the FP determinant and will be ignored in what follows since its existence
does not change our results.

We wish to get the Einstein-Hilbert term from ${\cal{L}}$ in (\ref{QG-Lag1}) plus ${\cal{L}}_{GF + FP}$
in (\ref{GF+FP2}). To do that, we first eliminate the $\tilde R^2$ term by choosing the gauge parameter $\alpha$
to be
\begin{eqnarray}
\alpha = \frac{1}{\xi_1}.
\label{gauge para1}
\end{eqnarray}
With this gauge parameter we obtain a quantum Lagrangian density:
\begin{eqnarray}
{\cal{L}}_q = \sqrt{-g} \left( \xi_1 k \tilde R - \frac{1}{4} H_{\mu\nu} H^{\mu\nu} - \frac{1}{2} \xi_1 k^2 \right).
\label{quant-Lag1}
\end{eqnarray}
Next, selecting the constant $k$ to be
\begin{eqnarray}
k = \frac{M_{Pl}^2}{2 \xi_1},
\label{k-choice}
\end{eqnarray}
where $M_{Pl}$ is the reduced Planck mass defined as $M_{Pl} = \frac{1}{\sqrt{ 8 \pi G}} = 
2.44 \times 10^{18} GeV$ and using (\ref{W-scalar-curv}), we arrive at the final expression up to a total 
derivative term: 
\begin{eqnarray}
{\cal{L}}_q = \sqrt{-g} \left[  \frac{M_{Pl}^2}{2} ( R - 2 \Lambda ) - \frac{1}{4} H_{\mu\nu} H^{\mu\nu} 
- \frac{1}{2} m_S ^2 S_\mu S^\mu \right],
\label{quant-Lag2}
\end{eqnarray}
where the cosmological constant $\Lambda$ and the mass of the Weyl gauge field, $m_S$, are positive 
definite and given by
\begin{eqnarray}
\Lambda = \frac{1}{8 \xi_1} M_{Pl}^2,  \qquad
m_S = \sqrt{6} f M_{Pl}.
\label{CC1}
\end{eqnarray}
Here it is of interest to notice that the present formalism provides only a positive cosmological constant,
which is very small at the weak coupling limit of the gravitational interaction, $\xi_1 \gg 1$. Moreover,
the mass of the Weyl gauge field depends on the coupling constant $f$ for the Abelian group.

In this way, starting with the classical Lagrangian density (\ref{QG-Lag1}), we have performed the BRST 
quantization for the Weyl gauge symmetry by taking the Weyl gauge (\ref{W-gauge}), and then succeeded in 
obtaining a quantum Lagrangian density (\ref{quant-Lag2}), which is sum of the Einstein-Hilbert term 
plus the massive Weyl gauge field, by choosing the gauge parameter Eq. (\ref{gauge para1}) and 
fix the constant $k = \frac{M_{Pl}^2}{2 \xi_1}$. As a result of this process, we have spontaneous symmetry
breakdown (SSB) of the Weyl gauge symmetry. However, it is unclear in the present gauge choice 
whether or not the number of dynamical degrees of freedom remains the same before and after the SSB, 
and what dynamical degree of freedom is eaten by the Weyl gauge field.

To answer these questions, it is convenient to move from the Jordan frame to the Einstein frame \cite{Fujii, Oda2}.  
First, let us rewrite the Lagrangian density (\ref{QG-Lag1}) by introducing a scalar field $\varphi$ of weight $-2$ 
as follows:
\begin{eqnarray}
{\cal{L}} = \sqrt{-g} \left( \varphi \tilde R - \frac{1}{2 \xi_1} \varphi^2 - \frac{1}{4} H_{\mu\nu} H^{\mu\nu} \right),
\label{QG-Lag2}
\end{eqnarray}
where the field equation for $\varphi$ reads
\begin{eqnarray}
\varphi = \xi_1 \tilde R.
\label{varphi-eq}
\end{eqnarray}
Substituting Eq. (\ref{varphi-eq}) into Eq. (\ref{QG-Lag2}) reproduces the classical Lagrangian density (\ref{QG-Lag1})
as required.

In order to move to the Einstein frame, let us first invert (\ref{Weyl transf 2}):
\begin{eqnarray}
g_{\mu\nu} = \Omega^{-2} (x) \hat g_{\mu\nu},
\label{Eins-frame}
\end{eqnarray}
where we have defined $\Omega (x) = e^{\Lambda (x)}$. Various geometrical quantities can be rewritten 
in terms of $\hat g_{\mu\nu}$ in a new conformal frame.
\begin{eqnarray}
g^{\mu\nu} &=& \Omega^{2} (x) \hat g^{\mu\nu}, \qquad \sqrt{- g} = \Omega^{-4} (x) \sqrt{- \hat g},
\nonumber\\
\Gamma^\lambda _{\mu\nu} &=& \hat \Gamma^\lambda _{\mu\nu} - ( F_\mu \delta^\lambda _\nu
+ F_\nu \delta^\lambda _\mu - \hat g^{\lambda\rho} F_\rho \hat g_{\mu\nu} ), 
\nonumber\\
R &=& \Omega^2 (x) ( \hat R + 6 \widehat{\Box} F - 6 \hat g^{\mu\nu} F_\mu F_\nu ),
\label{Quant-Eins-frame}
\end{eqnarray}
where we have defined
\begin{eqnarray}
F &=& \log \Omega, \qquad F_\mu = \partial_\mu F = \frac{\partial_\mu \Omega}{\Omega}, 
\nonumber\\
\widehat{\Box} F &=& \frac{1}{\sqrt{ - \hat g}} \partial_\mu \left( \sqrt{- \hat g} \hat g^{\mu\nu} 
\partial_\nu F \right).
\label{Quant-Eins-frame2}
\end{eqnarray}

Using these expressions, one can rewrite the Lagrangian density (\ref{QG-Lag2}) into a form
\begin{eqnarray}
{\cal{L}} &=& \sqrt{- \hat g} \biggl[ \varphi \Omega^{-2} \biggl( \hat R - 6 f \hat \nabla_\mu S^\mu 
- 6 f^2 \hat g^{\mu\nu} S_\mu S_\nu + 6 \widehat \Box F - 6 \hat g^{\mu\nu} F_\mu F_\nu
\nonumber\\
&+& 12 f \hat g^{\mu\nu} F_\mu S_\nu \biggr) - \frac{1}{2 \xi_1} \Omega^{-4} \varphi^2 
- \frac{1}{4} \hat g^{\mu\nu} \hat g^{\alpha\beta} H_{\mu\alpha} H_{\nu\beta} \biggr].
\label{QG-Lag3}
\end{eqnarray}
To reach the Einstein frame, we choose the scale factor $\Omega (x)$ such that
\begin{eqnarray}
\varphi \Omega^{-2} = \frac{1}{2} M^2 _{Pl}.
\label{Omega}
\end{eqnarray}
Then, up to total derivative terms, the Lagrangian density (\ref{QG-Lag3}) is cast to 
an expression
\begin{eqnarray}
{\cal{L}} &=& \sqrt{- \hat g} \biggl( \frac{M^2 _{Pl}}{2} \hat R 
- 3 f^2 M^2 _{Pl} \hat g^{\mu\nu} \hat S_\mu \hat S_\nu 
- \frac{1}{8 \xi_1} M^2 _{Pl} 
\nonumber\\
&-& \frac{1}{4} \hat g^{\mu\nu} \hat g^{\alpha\beta} H_{\mu\alpha} H_{\nu\beta} \biggr),
\label{QG-Lag4}
\end{eqnarray}
where we have defined
\begin{eqnarray}
\hat S_\mu = S_\mu - \frac{1}{2 f} \partial_\mu \log \varphi.
\label{hat-S}
\end{eqnarray}
Note that this Lagrangian density  (\ref{QG-Lag4}) coincides with the previous one (\ref{quant-Lag2}) with
(\ref{CC1}) (up to the FP ghost term).  

The point of the above derivation is twofold. First, the Weyl gauge field $S_\mu$ ``eats'' a scalar 
field $\varphi$ as seen in Eq. (\ref{hat-S}), thereby the Weyl gauge field becoming massive, and the number of 
dynamical degrees of freedom remains the same before and after the SSB as expected. 
In other words, a scalar mode hidden in the curvature squared is absorbed into the longitudinal mode
of the massless Weyl gauge field, by which the Weyl gauge field becomes massive. Second, the condition  
(\ref{Omega}) corresponds to the ``Einstein gauge'' or ``unitary gauge'' for which the scalar field takes 
a constant value. Indeed, since the scalar field $\varphi$ has weight $-2$, Eq.  (\ref{Omega}) means that  
in a new conformal frame, i.e., the Einstein frame, the scalar field takes a constant
\begin{eqnarray}
\varphi \rightarrow \hat \varphi = \Omega^{-2} (x) \varphi = \frac{1}{2} M^2 _{Pl}.
\label{U-gauge}
\end{eqnarray}
Incidentally, it turns out that the values of the cosmological constant and the mass of the Weyl gauge field in (\ref{CC1}) 
are independent of the gauge choice so that they are physical quantities.

\section{A more general gauge and spontaneous symmetry breakdown}

Encouraged by the observation in the previous section that spontaneous symmetry breakdown of the Weyl gauge
symmetry occurs, we will search for the $\it{genuine}$ spontaneous symmetry breakdown in the sense that 
a potential term determines the vacuum expectation value among arbitrary configurations.    

Before doing so, it is worthwhile to summarize the SSB found in the previous section and point out its problem. 
In the latter part of the previous section, we have started with a quadratic gravity (\ref{QG-Lag1}) in the Weyl geometry 
and then replaced it with a scalar-tensor gravity (\ref{QG-Lag2}) which is quantum mechanically equivalent to 
the quadratic gravity (\ref{QG-Lag1}). In the process of moving from the Jordan frame to the Einstein frame,
in Eq.  (\ref{Omega}) we have chosen a dimensional constant, i.e., the Planck mass $M_{Pl}$, to compensate for 
the dimension of mass squared of the scalar field $\varphi$. This change of the frames is found to be equivalent to 
taking the Einstein or unitary gauge. The introduction of the Planck mass into a Weyl invariant theory has triggered 
the SSB of the Weyl gauge symmetry. To put differently, the choice of the gauge condition has given rise to the SSB.
This situation also holds in the arguments of the former part of the previous section where the choice of the Weyl
gauge provided us with the SSB of the Weyl gauge symmetry.    

Let us recall that in the conventional scenario of the SSB, there exists a potential inducing the SSB while we have 
no such potential in the theory under consideration. Nevertheless, the very presence of a solution with dimensional constants
or the gauge condition for the Weyl gauge symmetry justifies the claim that the present scenario of the SSB is 
nothing but spontaneous symmetry breakdown. Actually, the SSB happens in such a way that one dynamical degree 
of freedom is ``eaten'' by a gauge field and as a result the gauge field becomes massive, and the number of dynamical
degrees of freedom remains unchanged before and after the SSB.
 
A possible problem, however, arises from the lack of a suitable potential in the sense that we cannot single out 
a solution realizing the SSB on the stability argument \cite{Fujii}. In order to overcome this problem, we have derived
an effective potential displaying the SSB from the radiative corrections of gravitational fields stemming from 
higher derivative terms \cite{Oda1, Oda2}\footnote{This study is based on our previous paper \cite{Oda3}.} by 
using the Coleman-Weinberg mechanism \cite{Coleman}. In this section, we would like to derive such a potential 
by using the gauge fixing condition. At first sight, it seems to be strange that one can derive a Higgs potential 
from the gauge fixing procedure since it is thought that the choice of a gauge fixing condition is at one's disposal 
and physics does not usually depend on such a gauge condition. However, in case of scale invariant gravitational
theories \cite{Oda4, Oda5, Oda6}, the gauge fixing of a local scale symmetry or the Weyl gauge symmetry is
usually associated with the introduction of dimensional constants (except for the Lorenz gauge condition),
which triggers the SSB of the symmetry. In addition to it, our present study might provide a bold conjecture of
the origin of the Higgs potential that the Higgs potential comes from a suitable gauge fixing for 
a local scale symmetry or the Weyl gauge symmetry. 
 
We are now ready to start with a more general theory whose action is given in Eq.  (\ref{Action2}). 
Since $\tilde R$ and $\phi^2$ are invariant under diffeomorphisms and have weight $-2$, one can construct
a gauge invariant line element:
\begin{eqnarray}
d \tilde s^2 &=& ( a \tilde R + b \phi^2 ) g_{\mu\nu} d x^\mu d x^\nu \nonumber\\
&\equiv& ( a \tilde R + b \phi^2 ) d s^2,
\label{line-element2}
\end{eqnarray}
where $a$ and $b$ are constants.  The requirement that this gauge invariant line element 
should be reduced to the line element in the Riemann geometry by taking a gauge condition gives us
information on a more general gauge condition 
\begin{eqnarray}
a \tilde R + b \phi^2 = k,
\label{O-gauge}
\end{eqnarray}
where $k$ is a certain constant to be determined shortly. 

Following the same line of the argument as before, let us use the BRST formalism to construct 
a quantum theory. In addition to the BRST transformation (\ref{BRST1}), we have a BRST
transformation for $\phi$; since the scalar field $\phi$ has weight $-1$, its BRST transformation 
is of form 
\begin{eqnarray}
\delta_B \phi = - c \phi.
\label{BRST2}
\end{eqnarray}
Then, the Lagrangian density for the gauge fixing term and the FP ghost term is given by
\begin{eqnarray}
{\cal{L}}_{GF + FP} &=& - i \delta_B \left[ \sqrt{- g} \bar c ( a \tilde R + b \phi^2 - k 
+ \frac{1}{2} \alpha B ) \right]
\nonumber\\
&=& - \frac{1}{2 \alpha} \sqrt{- g} ( a \tilde R + b \phi^2 - k )^2 - i \hbar \delta^4 (0) 
\log \sqrt{- g(x)}.
\label{O-GF+FP}
\end{eqnarray}
where we have performed the integration over $B_*, c, \bar c$ as before.
The last term again comes from the FP determinant and will be ignored below.

Adding the classical Lagrangian density in (\ref{Action2}), and the gauge-fixing and FP ghost
Lagrangian density (\ref{O-GF+FP}), we have 
\begin{eqnarray}
{\cal{L}}_q &=& \sqrt{-g} \Biggl\{  \frac{1}{2} \left( \xi_1 - \frac{a^2}{\alpha} \right) \tilde R^2 
+ \left[ \left( \frac{\xi_2}{2} - \frac{a b}{\alpha} \right) \phi^2 + \frac{a k}{\alpha} \right] \tilde R
\nonumber\\
&-& \frac{1}{4} H_{\mu\nu} H^{\mu\nu}  - \frac{1}{2} g^{\mu\nu} D_\mu \phi D_\nu \phi 
- \frac{1}{2 \alpha} ( b \phi^2 - k )^2 - \frac{\lambda}{4} \phi^4 \Biggr\}.
\label{O-quant-Lag1}
\end{eqnarray}
To remove the term involving $\tilde R^2$, we shall take the gauge parameter $\alpha$
to be
\begin{eqnarray}
\alpha = \frac{a^2}{\xi_1}.
\label{gauge para2}
\end{eqnarray}
With this choice of the gauge parameter,  the Lagrangian density (\ref{O-quant-Lag1}) becomes
\begin{eqnarray}
{\cal{L}}_q &=& \sqrt{-g} \Biggl\{  \left[ \left( \frac{\xi_2}{2} - \frac{\xi_1 b}{a} \right) \phi^2 
+ \frac{\xi_1 k}{a} \right] \tilde R - \frac{1}{4} H_{\mu\nu} H^{\mu\nu} 
\nonumber\\
&-& \frac{1}{2} g^{\mu\nu} D_\mu \phi D_\nu \phi - \frac{\xi_1}{2 a^2} ( b \phi^2 - k )^2
- \frac{\lambda}{4} \phi^4 \Biggr\}
\nonumber\\
&=& \sqrt{-g} \Biggl\{  \left[ \left( \frac{\xi_2}{2} - \frac{\xi_1 b}{a} \right) \phi^2 
+ \frac{\xi_1 k}{a} \right] \tilde R - \frac{1}{4} H_{\mu\nu} H^{\mu\nu} 
\nonumber\\
&-& \frac{1}{2} g^{\mu\nu} D_\mu \phi D_\nu \phi 
- \frac{\xi_1}{2 a^2} \left( b^2 + \frac{a^2 \lambda}{2 \xi_1} \right)
\left( \phi^2 - \frac{b k}{b^2 + \frac{a^2 \lambda}{2 \xi_1}} \right)^2
\nonumber\\
&-& \frac{k^2 \lambda}{4 \left( b^2 + \frac{a^2 \lambda}{2 \xi_1} \right)} \Biggr\}.
\label{O-quant-Lag2}
\end{eqnarray}
It is worthwhile to notice that the Higgs potential has emerged in the Lagrangian density from the
gauge fixing condition (\ref{O-gauge}) for the Weyl gauge transformation due to $\xi_1 > 0, \lambda > 0$.
The present theory, therefore, solves the ``possible' problem'' mentioned above and the presence
of the Higgs potential makes it possible to single out a solution realizing the SSB on the stability 
argument. 

Taking the vacuum expectation value 
\begin{eqnarray}
\langle \phi^2 \rangle = \frac{b k}{b^2 + \frac{a^2 \lambda}{2 \xi_1}},
\label{VEV}
\end{eqnarray}
we have a quantum Lagrangian density in the lowest level of approximation
\begin{eqnarray}
{\cal{L}}_q &=& \sqrt{-g} \Biggl[ \frac{k ( \xi_2 b + a \lambda )}{2 (b^2 + \frac{a^2 \lambda}{2 \xi_1})} \tilde R 
- \frac{1}{4} H_{\mu\nu} H^{\mu\nu} 
\nonumber\\
&-& \frac{1}{2} \frac{f^2 b k}{b^2 + \frac{a^2 \lambda}{2 \xi_1}} S_\mu S^\mu 
- \frac{k^2 \lambda}{4 \left( b^2 + \frac{a^2 \lambda}{2 \xi_1} \right)} \Biggr].
\label{O-quant-Lag3}
\end{eqnarray}
To make this Lagrangian density coincide with the Einstein-Hilbert Lagrangian, we choose the constant
$k$ to be
\begin{eqnarray}
k = \frac{b^2 + \frac{a^2 \lambda}{2 \xi_1}}{\xi_2 b + a \lambda} M^2 _{Pl}.
\label{k-choice2}
\end{eqnarray}
As a result, up to a total derivative term, (\ref{O-quant-Lag3}) takes the form
\begin{eqnarray}
{\cal{L}}_q &=& \sqrt{-g} \Biggl[  \frac{M_{Pl}^2}{2} R - \frac{1}{4} H_{\mu\nu} H^{\mu\nu} 
- \frac{1}{2} \frac{( 6 \xi_2 + 1) b + 6 a \lambda}{\xi_2 b + a \lambda} ( f M_{Pl} )^2 S_\mu S^\mu 
\nonumber\\
&-& \frac{\lambda \left( b^2 + \frac{a^2 \lambda}{2 \xi_1} \right)}{4 (\xi_2 b + a \lambda)^2} M^4 _{Pl}  \Biggr].
\label{O-quant-Lag4}
\end{eqnarray}
From this Lagrangian density, one can easily read out the values of the cosmological constant and
the mass of the Weyl gauge field as
\begin{eqnarray}
\Lambda &=& \frac{\lambda \left( b^2 + \frac{a^2 \lambda}{2 \xi_1} \right)}{4 (\xi_2 b + a \lambda)^2} M^2 _{Pl},
\nonumber\\ 
m_S &=& \sqrt{\frac{( 6 \xi_2 + 1) b + 6 a \lambda}{\xi_2 b + a \lambda}} f M_{Pl}.
\label{CC2}
\end{eqnarray}
Note that the the cosmological constant is again positive definite and very tiny in the weak coupling regime
$\xi_2 \gg 1, \lambda \ll 1$.\footnote{Even if the cosmological constant is tiny in the weak coupling limit,
this is no longer a solution to the cosmological constant problem as proved in Refs. \cite{Weinberg, Oda7, Oda8,
Oda9}.} Of course, the number of dynamical degrees of freedom is unchanged before and after the SSB,
and a scalar field in the scalar curvature squared is eaten by the massless Weyl gauge field as in the previous
model.

\section{Conclusions}
In this article, we have investigated a possibility that gravitational theories in the Weyl geometry, which
are invariant under the Weyl gauge transformation as well as diffeomorphisms, generate the Einstein-Hilbert
action of Einstein's general relativity. We have seen that a quadratic gravity without a scalar field induces
the Einstein-Hilbert action and at the same time the Weyl gauge field becomes massive in the ``Weyl gauge'' 
where Weyl's scalar curvature takes a constant value. Furthermore, we have showed that the same phenomenon 
also occurs in the ``Einstein gauge'' or ``unitary gauge'' for which the scalar field takes a constant value.
The latter gauge choice clarifies that a scalar field included in the scalar curvature squared is absorbed into
a longitudinal mode of the massless Weyl gauge field, thereby the gauge field becoming massive. We also
pointed out that this phenomenon is nothing but spontaneous symmetry breakdown (SSB) \cite{Fujii, Oda2} in the
sense that the gauge field becomes massive and the number of physical degrees of freedom remains unchanged
before and after the SSB, but the absence of the Higgs potential provides us a problem that one cannot single
out a solution realizing the SSB on the stability argument.
  
In order to solve the problem of the absence of the Higgs potential, we have looked for a more general model 
of a quadratic gravity with a scalar field in the Weyl geometry. This general model in fact shows an interesting 
phenomenon that a more general gauge fixing condition yields a Higgs potential whose minimum determines 
the vacuum expectation value, thereby triggering spontaneous symmetry breakdown. In this model,
the Einstein-Hilbert term is also induced and the Weyl gauge field becomes massive via the SSB.

The key observation behind the present work is to overcome the second clock problem associated with
the Weyl geometry. The second clock problem says that two synchronized clocks put at a certain space-time point
run at different speeds after they travel along different paths and then reunite at some place. To put
differently, two identical elementary particles with the same mass show different masses after they get
the same experience as above since mass is nothing but length scale in this context. However, in this
interpretation, there is an implicit assumption such that the length $l$ has a variation
\begin{eqnarray}
d l = f l S_\mu d x^\mu,
\label{d l}
\end{eqnarray}
and its integration $l$ given by    
\begin{eqnarray}
l = l_0 e^{f  \int S_\mu d x^\mu},
\label{l}
\end{eqnarray}
determines a physcal length. As long as we admit this implicit assumption about the physical length,
we are forced to encounter the second clock problem. One famous resolution, which eventually led
to the concept of the $U(1)$ gauge invariance, is to replace the scale factor $f$ with a imaginary phase 
$i f$. Then, the second clock problem amounts to the question of whether, say, two electrons starting 
with the same phase and taken along different paths would acquire different phases or not. The experimental 
answer is definitely yes and this phenomenon is nothing but the Aharonov-Bohm effect. 

An alternative resolution, which we adopt in this article, is that we do not regard 
$d s^2 = g_{\mu\nu} d x^\mu d x^\nu$ or $l$ as a physical quantity measuring the distance since it is 
not gauge invariant. The gauge invariant line element $d \tilde s$ defined in Eq. (\ref{line-element1}) or
Eq. (\ref{line-element2}) should be regarded as the definition of the
distance in the Weyl geometry.  It is then obvious that we are free from the second clock problem.   
If we require that this gauge invariant line element $d \tilde s$ is reduced to $d s$ in the Riemann
geometry after fixing the Weyl gauge invariance, we are led to taking the gauge condition, the
Weyl gauge or the more general gauge.   
   
To close this article, let us comment on the Higgs potential. Let us recall that there is a longstanding
and important question of what the origin of the Higgs potential is. Our model might partially answer 
this question: The Higgs potential comes from symmetry breaking of a local scale symmetry or Weyl gauge 
symmetry. In this context, it is worth recalling that the standard model is invariant under a scale symmetry 
if the (negative) mass term for the Higgs field is removed from its action \cite{Bardeen}.  In this sense,
it is naturally expected that the appearance of the Higgs potential is closely related to the scale
symmetry. The Weyl geometry provides us with a playground for promoting a $\it{global}$ scale symmetry
to a $\it{local}$ scale symmetry and simultaneously produces a Weyl gauge field as a candidate for dark
matter.


\end{document}